\documentclass[12pt]{amsart}
\usepackage[utf8]{inputenc} 
\usepackage{graphicx}
\usepackage{amsmath}
\usepackage[usenames, dvipsnames]{color}
\usepackage{eso-pic}

\def\tmp#1{}

\def\be{\begin{equation}}
\def\ee{\end{equation}}

\newcommand{\avwap}{$\alpha$VWAP{}}
\newcommand{\apov}{$\alpha$POV{}}
\newcommand{\ais}{$\alpha$IS{}}

\newcommand{\tgt}{\text{tgt}}
\newcommand{\opt}{\text{opt}}

\begin{document}


\vskip.5cm
\begin{center}
{\large \bf Design and Implementation of Schedule-Based Trading Strategies Based on Uncertainty Bands\footnote{The Journal of Trading, Fall 2011, Vol. 6, No. 4, pp. 45-52}}\\

\vskip.1cm
\end{center}
\vskip0.5cm

\begin{center}
{\bf
{Vladimir Markov}$^{a}$, {Slava Mazur}$^{b}$
{\rm, and}
{David Saltz}$^{c}$}
\end{center}
\vskip 8pt

\begin{center}

$^{a}${\it vmarkov@liquidnet.com}\, ,
$^{b}${\it smazur@liquidnet.com}\, ,
$^{c}${\it dsaltz@liquidnet.com}\\
\vspace*{0.3cm}
\mbox{Liquidnet, 498 Seventh Avenue, New York, NY, 10018}\\
\end{center}

\vglue 0.3truecm



\section*{Introduction}

We propose a design for schedule-based trading strategies based on uncertainty bands.  This formulation (1) simplifies strategy specification and implementation; (2) provides for flexible allocation among passive, opportunistic, aggressive, and dark pool crossing execution tactics; (3) allows for rapid enhancements as new optimization methods, scheduling techniques, alpha models, and execution tactics are developed; and (4) yields information at macroscopic (strategic) and microscopic (tactical) levels that is easily published to trading databases and front-end applications.

Although there is an abundance of literature on optimal trading strategies  (Kissell and Glantz [2003]), the practical implementation is usually comprised of heuristic rules which are necessary to incorporate and parameterize ``stylized'' facts of the financial markets (Bouchaud  and Potters [2004]), handle edge cases and other departures of the implementation from the theoretical model, and maintain flexibility to satisfy client needs.  Many schedule-based trading algorithms utilize a single target trajectory with an aggressiveness model that takes into account the current executed shares in comparison to the target and the current and recent market conditions. The popularity of this approach is explained by its simplicity of implementation.  A major deficiency is the entanglement of the high-level strategy that defines the macroscopic schedule and the execution tactics that efficiently capture bid-ask spread and minimize immediate market impact and adverse price selection of passive orders. The absence of tactics encapsulation makes it difficult to re-use low-level trading logic and to back-test individual components of the trading algorithms.

Our proposal is to implement schedule-based strategies in the framework of {\em uncertainty bands}, which are trading trajectories that define the outer limits of order slicing behavior.  These bands represent the uncertainty of trade scheduling in noisy markets, the discretion provided by the client to the strategy, or any combination thereof.  Just as a confidence interval provides more insight into a probability distribution than a single mean value, uncertainty bands provide more insight into the potential paths of order execution than a single target trajectory.

\section*{Formulation of the Approach}

The primary examples of schedule-based strategies are Participation (Percentage of Volume, or POV), Volume-Weighted Average Price (VWAP), and Implementation Shortfall (IS).  All such strategies work an order of $X_0$ shares over a target trajectory $X_\tgt{}(t)$, where $t$ is time.  The order begins trading at $t=t_0$ and it ceases trading at $t=t_1$.  Sometimes $t_1$ is known and sometimes $X_\tgt(t_1) = X_0$, {\em i.e.}, the order is expected to be completely filled, but neither condition is guaranteed.  Moreover, in some cases $X_\tgt(t)$ is known in advance for all $t > t_0$ but typically it is not.  In a real-time trading strategy, it is sufficient that $X_\tgt(t)$ be known only when $t$ is the current clock time.  In all cases, $X_\tgt(t) = 0$ for all $t < t_0$.

We refer to $X_\tgt(t)$ as the {\em schedule target}.  The strategy attempts to realize a trajectory $X_f(t)$ that closely approximates the schedule target.  To the extent that $X_f(t)$ can depart from $X_\tgt(t)$, the strategy has more discretion to seek price improvement.  A large departure of $X_f(t)$ from $X_\tgt(t)$ can produce a negative client experience and a higher performance risk.  The amount of available discretion is a function of the strategy model, the stock's historical trading characteristics, the market conditions, and the client's trading instructions such as maximum participation rate or risk aversion.

Define $X_\text{max}(t)$ to be the upper trajectory, or upper {\em uncertainty band}, so that the strategy obeys the constraint $0 \le X_f(t) \le X_\text{max}(t) \le X_0$ for all $t$.  Define $X_\text{min}(t)$ to be the lower uncertainty band, so that the strategy obeys the constraint $X_0 \ge X_f(t) \ge X_\text{min}(t) \ge 0$ for all $t$.  In practice it is impossible to enforce these constraints 100\% of the time.  Instead, the strategy acts so as to make $X_f > X_\text{max}$ unlikely, and when it does occur the strategy pauses trading in displayed (or perhaps all) venues until the constraint is satisfied.  Whenever $X_f < X_\text{min}$, the strategy executes the shortfall $X_\text{min} - X_f$ aggressively in displayed markets, paying the spread and any resulting impact costs, so as to minimize the time during which the constraint is violated.

The child orders (slices) that are sent by the strategy to implement the schedule are, at a high level, comprised of market and limit orders.  Market orders are aggressively priced orders routed to displayed markets that are used to quickly cover any shortfall
\be
X_A(t) = \max\{0, X_\text{min}(t) - X_f(t) \} ~.
\ee
Market orders may also be used in discretionary situations for opportunistic liquidity capture.  Limit orders are all other orders, priced to capture as much of the bid-ask spread as possible while minimizing market impact.  Note that market orders do not necessarily have the market order type; in most cases they are orders with a limit order type whose price is set aggressively enough to execute immediately (buy at the ask price, sell at the bid price).  Market and limit orders in this context are also referred to as aggressive and passive orders, respectively.  Within the class of limit orders are dark (ATS) orders, which might be priced aggressively but typically execute at a price within the spread such as the midpoint.

The residual shares available for passive or opportunistic slicing is
\be
X_P(t) = \max\left\{ 0, X_\text{max}(t) - \max\{X_f(t), X_\text{min}(t) \} \right\} ~.
\ee
These shares represent the discretion available to the strategy to seek spread and opportunistic liquidity capture.  Typically these shares are split into child orders of varying price levels that are routed to multiple execution venues.  The precise manner in which this routing is accomplished is in the domain of {\em market execution tactics}.  The strategy engine should allow maximum flexibility to select a combination of execution tactics based on a variety of factors, including order attributes, client preferences, and market conditions. Depending on its function, a tactic might be implemented as a smart order router (SOR) or as another strategy.  A very simple execution tactic is to keep the entire discretionary quantity $X_P$ posted at the best bid price (if buying; best ask price if selling) in the order book of any major displayed market.

Implementation Shortfall strategies attempt to fill as much of the client order as possible in dark pools that are considered to have negligible information leakage.  Executions in such dark pools do not incur market impact and have zero bid-ask spread costs.  Still, protective measures have to be taken as dark pool orders can be adversely selected by more informed traders (Altunata,  Rakhlin, and Waelbroeck [2010]).  At any time $t$, the number of shares available for execution in dark pools is
\be
X_D(t) = X_0 - X_\text{max}(t) ~,
\ee
in addition to whatever portion of $X_P(t)$ is also made available to the pools as determined by the market execution tactics.  In the event of a (typically block-sized) dark execution, the size of the execution is immediately added to all of the trajectories $X_\text{min}$, $X_\text{max}$, $X_\tgt$, and $X_f$.

Participation in dark pools is managed by any combination of the available {\em dark pool tactics}, analogous to the handling of limit orders.  The strategy engine should allow maximum flexibility to select a combination of dark pool tactics based on a variety of factors, including order attributes, client preferences, and market conditions (Glukhov [2007]).  A very simple dark pool tactic is to route the entire amount $X_D(t)$ to the Liquidnet $\text{H}_2\text{O}$ ATS.  Besides IS strategies, POV strategies that are calibrated to the volume rates in displayed markets also attempt to cross as many shares as possible in dark pools.

This formulation clearly distinguishes between the strategy's high-level scheduling and its low-level tactical execution components.  At the tactical level, the precise manner in which the trajectories are computed is irrelevant, thus a single tactical driver serves the needs of all schedule-based strategies.

Finally, note that the bands $X_\text{min}$ and $X_\text{max}$ are used to partition the order residual into a three-way allocation of active, passive, and dark shares.  Alternatively, the schedule target $X_\text{tgt}$ can be introduced to further split the passive shares, {\em e.g.},
\begin{eqnarray*}
X_{P_1}(t) & = & \max\left\{ 0, X_\text{tgt}(t) - \max \{ X_f(t), X_\text{min}(t) \} \right\} ~, \\
X_{P_2}(t) & = & X_P(t) - X_{P_1}(t) ~.
\end{eqnarray*}
The quantity $X_{P_1}(t)$ represents an additional shortfall with respect to the target trajectory that is not covered by the aggressive shares.  If the current market conditions are considered to be favorable for market execution (as indicated by, for example, an $\alpha$ model) then one might execute $X_{P_1}(t)$ aggressively, leaving only $X_{P_2}(t)$ for passive market participation.

\section*{Implementation: Continuous-Time Approach}
Below we describe reference implementations of the schedule-based strategies VWAP, POV, and IS.  We call these implementations \avwap, \apov, and \ais, respectively, because of the simplicity of incorporation of alpha models into this framework.  These models are presented to demonstrate the practicality of our approach and should not be construed as the models underlying Liquidnet's trading strategies.

\subsection*{VWAP}
\begin{figure}[h]
\centering
\includegraphics[width=3in]{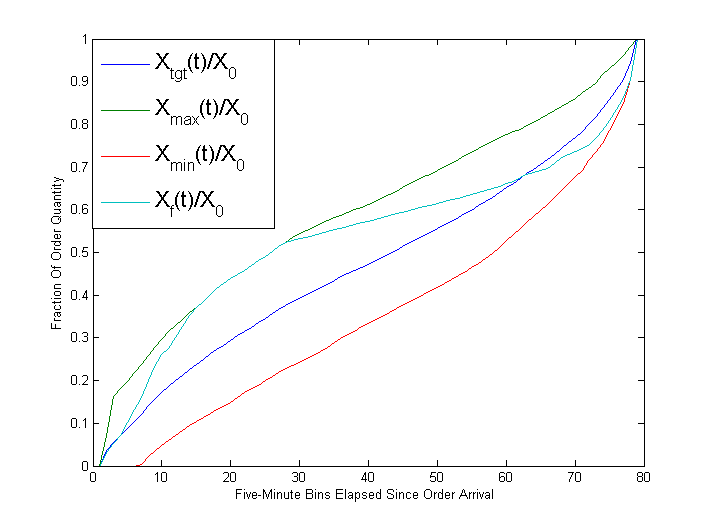}
\caption{Example of \avwap\ trading trajectory with uncertainty bands}
\label{fig:States}
\end{figure}

The conventional VWAP strategy follows a historical intraday volume distribution, while the \avwap\ strategy allows a reasonable deviation from the historical curve to capture price and liquidity opportunities, encapsulated in the bands $X_\text{min}(t)$ and $X_\text{max}(t)$. Define the volume curve $U(t)$ to be the fraction of the stock's daytime trading volume that is executed as of time $t$. As the volume curve has an inherent uncertainty, a banded schedule built around a confidence interval for the random variable $U(t)$ is appropriate. Let $u(t)$ be the volume curve normalized and bounded for our trading schedule,
\be
u(t) = \frac{U(t) - U(t_0)}{U(t_1) - U(t_0)} ~,
\ee
and let $\bar{u}(t)$ and $\delta u(t)$ be, respectively, the mean and standard deviation of $u(t)$.  The first and second moments of $U(t)$ can be determined from any combination of empirical data (historical with possibly intraday adjustments) and a model probability distribution.  The schedule target is
\be
X_\tgt(t) = \bar{u}(t) X_0 ~.
\ee
If $u(t)$ is assumed to be symmetrically distributed around its mean, then we form a confidence interval around the schedule target whose width is measured in standard deviation units:
\begin{eqnarray*}
X_\text{min}(t)  & = & \max \{ 0,     X_\tgt(t) - \eta \delta u(t) X_0 \} ~, \\
X_\text{max}(t) & = & \min  \{ X_0, X_\tgt(t) + \eta \delta u(t) X_0 \} ~.
\end{eqnarray*}
The dimensionless parameter $\eta$ represents the discretion afforded to the strategy to depart from the schedule target.   Examples of trading trajectories calculated in this model are shown in Figure \ref{fig:States}.  Also shown is the realized trajectory $X_f(t)$ which demonstrates how the order is successfully completed within the uncertainty bands.

More realistically, the distribution of $u(t)$ is asymmetric, so a band structure based on quantiles is appropriate (Mazur [2011]) :
\begin{eqnarray*}
X_\text{max}(t) & : & P\left\{u(t)X_0 > X_\text{max}(t)\right\} = q ~, \\
X_\text{min}(t) & : & P\left\{u(t)X_0 < X_\text{min}(t)\right\} = q ~,
\end{eqnarray*}
where $q: 0 < q < 1$ is the discretion parameter, namely the $100(1-q)\%$ confidence level of the uncertainty bands.  Small $q$ corresponds to a large deviation from the historical trajectory $X_\tgt(t)$, thus more discretion to the strategy.

As a strategy implemented in the uncertainty band framework, \avwap\ works a portion $X_A$ of the parent order actively and another portion $X_P$ passively.  The active (shortfall) component is executed with aggressively priced orders and ensures that the realized trajectory $X_f(t)$ is bounded below by $X_\text{min}(t)$.  The passive component is exposed to spread and liquidity capture opportunities and is limited in size so as to guarantee that $X_f(t)$ is bounded above by $X_\text{max}(t)$.  In a ``strict'' VWAP strategy, the residual shares $X_D$ are not available for crossing, but some of the passive shares should be exposed to dark venues.

A short-term price prediction ($\alpha$) model is used within the uncertainty band framework to exploit stock-specific price patterns such as whether a mean reversion or trend continuation is expected in the wake of a short-term price spike.  The dominant behavior can be estimated, for example, by analyzing the expected value of a future price increment (``response") conditioned on the value of a realized price increment (``push") at various timescales (Zaitsev, Zaitsev, Leonidov, and Trainin [2009]).  When executing a buy order subject to a strong short-term buy signal, a portion of $X_P$ that is normally executed passively is instead aggressively priced.  The size of this portion scales with the strength of the signal. Inversely, a strong sell signal triggers logic that pulls a portion of $X_P$ (and possibly $X_D$) away from the market. Besides the price improvement, the probability of the fill of a limit order increases and the market impact of a market order decreases in mean-reversion models as the algorithm trades against the short-term trend. Profit opportunities less than the the transaction costs  are not relevant for buy-side firms but represent significant interest for sell-side firms.

\subsection*{Participation}
Another example of a concrete strategy implemented in the framework of uncertainty bands is \apov.
The calculation of the lower, upper, and target trajectories is straightforward given the client's respective minimum, maximum, and target participation rates.  In practice, the client might specify a single, target participation rate $p_\tgt$ with a tolerance that implies the range $\left[p_\text{min}, p_\text{max}\right]$, or the range is specified and $p_\tgt = \frac{1}{2}(p_\text{min} + p_\text{max})$ if it is not specified by the client.

Let $V_e(t)$ be the eligible volume.  The exact definition of eligible volume has some special cases but for the most part $V_e(t)$ is the volume traded on the order books of the displayed markets within the client's limit and during the interval $[t_0, t]$.  Assume that the participation rates are functions of time to allow for the possibility that they are modified by the client or set strategically by another algorithm. The lower trajectory is then
\be
X_\text{min}(t) = \int_{t_0}^t p_\text{min}(s) \dot{V}_e(s) ds ~,
\ee
and similarly for $X_\text{max}$ and $X_\text{tgt}$, where $\dot{V_e}$ denotes differentiation of $V_e$ with respect to its time argument.  In \apov, the same tactical driver that is used for \avwap\ distributes the active and passive shares among the market execution tactics, informed by an $\alpha$ model.

Similar considerations of ``strictness'' apply to POV.  The default behavior of a POV strategy is typically to maximize crossing in dark pools, thus the residual $X_D$ is fully allocated.  However, some clients are willing to forgo block crossing opportunities in order to spread the order execution over a longer time period, in which case the $X_D$ shares are not available for crossing.  If in the default version of POV a block trade of size $X_B$ is executed at time $t_B$, this block is not counted in the eligible volume but it is added to all of the trajectories as of time $t_B$, {\em e.g.},
\be
X_\text{tgt}(t) = \int_{t_0}^t p_\text{tgt}(s) \dot{V}_e(s) ds + X_B
\ee
for $t \ge t_B$.

\subsection*{Implementation Shortfall}
An Implementation Shortfall strategy can be designed as an adaptive POV strategy or in a mean-variance framework (Almgren and Chriss [2000]). In the adaptive POV approach, the strategy makes real-time adjustments to the participation rates $p_\text{min}(t)$, $p_\text{max}(t)$, and $p_\text{tgt}(t)$ in response to the trading environment, for example by increasing the participation rates as the market price moves in the client's favor or when substantial displayed liquidity materializes.  In the mean-variance approach the optimal trajectory is determined {\em a priori} by balancing forecasts of market impact and timing risk.  Adaptation to changes in the trading environment is accomplished by some combination of re-optimization and heuristic, real-time adjustments to the trading schedule.

Here we propose the \ais\ strategy as a practical implementation of the mean-variance approach.  The optimal trajectory represents a trade-off between expected market impact $I$ and timing risk $R$.  It is convenient to work with the residual shares $Y(t) = X_0 - X(t)$ and volume time $t$.  We model $Y(t)$ as a power-law trajectory over a volume duration $T$,
\be
Y(t) = X_0 \left( 1 - \frac{t}{T}\right)^\nu ~.
\label{eq:Y}
\ee
The optimal trajectory $Y_\opt(t)$ is parameterized by an optimal duration $T_\opt$ and shape parameter $\nu_\opt$.    With time in volume units, Eq.~(\ref{eq:Y}) with $\nu = 1$ is a VWAP trading schedule.

The estimated trading cost taking into account the client's aversion to timing risk is
\be
C[Y] = I[Y] + \frac{1}{2\rho} R^2[Y] ~,
\ee
where $\rho=(\sigma_D X_0 P_0) / A$, $\sigma_D$ is the daily return volatility, $P_0$ is the current stock price, and $A$ is a dimensionless risk aversion parameter.  With instantaneous market impact $J(\dot {Y}(s))$ and impact decay kernel $G(t - s)$ the expected cost of implementing a trading schedule $Y(t)$ is given by
\be
I[Y]=\int_0^T dt \dot {Y}(t) \int_0^t ds J(\dot {Y}(s)) G(t-s) ~.
\ee
The non-trivial decay kernel $G(t - s)$ accounts for the long-memory autocorrelation of trade signs in empirical high-frequency price data and has a power law behavior in general  (Gatheral [2010]).  Here, we assume that a trading tactic has a reasonable delay between trades to avoid amplification of the market impact through the decay kernel, so that $G(t-s) = \delta(t-s)$.

Empirical data suggest a power-law ansatz for the instantaneous market impact $J(\dot{Y})$ (Bouchaud, Farmer, and Lillo [2008])
\be
J(\dot{Y})=I_0 \sigma_D P_0 \left(\frac{\dot{Y}}{V_D}\right)^\beta ~,
\label{eq:instantaneous_impact}
\ee
where $V_D$ is the expected daily volume, $P_0$ is the price, and $I_0$ is a stock-dependent scale quantity.  Using this model and assuming an instantaneous decay kernel, we derive the expected implementation shortfall as a function of volume duration $T$ and parameterized by $\nu$ and $\beta$,
\be
I(T; \nu,\beta) = \left[ \frac{\nu^{\beta+1}}{1+(\nu-1)(\beta+1)}\right]
I_0 \sigma_D  X_0 P_0 \left(\frac{X_0}{T V_D}\right)^\beta
 ~.
\ee
The timing risk of the power-law trajectory is approximately ~  (Kissell and Glantz [2003])
\be
R^2(T; \nu) = \sigma_D^2 P_0^2 \int_0^T dt Y^2(t) = \frac{1}{(2\nu+1)} \sigma_D^2 X_0^2 P_0^2 T ~.
\ee
The optimal trading horizon $T_\opt$ is the value of $T$ that minimizes
\[
C(T; \nu, \beta) = I(T; \nu, \beta) + \frac{1}{2 \rho} R^2(T; \nu) ~.
\]
For simplicity, we set $\nu=1$ and obtain $T_\opt$ as the solution of $dC/dT = 0$,
\be
T_\opt = \left[ \frac{ 2 \beta \rho I(1; 1, \beta)}{ R^2(1; 1)} \right]^{\frac{1}{\beta+1}}=
\left(\frac{6 \beta I_0 }{A} \right)^{\frac{1}{\beta+1}} \left(\frac{X_0} {V_D } \right)^{\frac{\beta}{\beta + 1}} ~.
\ee
Note that a volume duration $T=1$ means exactly one trading day. Although the optimal trading time does not depend on volatility, the trading schedule does through the shape parameter $\nu$. The average participation rate corresponding to this schedule is
\be
p_\opt = \frac{X_0}{T_\opt V_D} = \left(\frac{A} {6 \beta I_0 } \right)^{\frac{1}{\beta+1}} \left(\frac{X_0} {V_D } \right)^{\frac{1}{\beta + 1}} ~.
\ee
The expected implementation shortfall of the optimal trading schedule is
\be
I(T_\opt) = {\left(\frac{A }{6 I_0 \beta}\right)^{\frac{\beta}{\beta+1}}}
\left[ \frac{\nu^{\beta+1}}{1+(\nu-1)(\beta+1)}\right] {I_0 \sigma_D P_0 X_0}
\left(\frac{X_0}{V_D}\right)^{\frac{\beta}{\beta+1}}
\sim \sigma_D X_0 \left(\frac{X_0}{V_D}\right)^{\frac{\beta}{\beta+1}}
\ee
The expected optimal market impact cost per share with empirically observed $\beta=0.5$  is $I_\opt /X_0 \sim \sigma_D \left(\frac{X_0}{V_D}\right)^{1/3}$.

The expected cost of trading $I_\text{pl}$ with the power law decay kernel $G_\text{pl}(t-s) = g_0 / |t - s|^\gamma$ is given by
\be
I_\text{pl}=T^{1-\gamma-\beta} I_0 \sigma_D g_0  X_0 P_0  \left (\frac{X_0}{V}\right)^{\beta} \nu^{\beta+1} \frac{ \Gamma (1-\gamma) \Gamma (\nu)} {(2-\gamma+(\beta+1)(\nu-1)) \Gamma(1-\gamma+\nu)} ~.
\ee
The coefficient $g_0$ can be estimated from historical execution data.  Empirically the exponent $\gamma\approx 0.5$  and $\gamma+\beta \approx 1$. The cost of naive continuous trading $I_\text{pl}$ is amplified by the decay kernel. It does not depend on the duration of trade for moderate rates of trading  and is given by the
"square-root" law $I_\text{pl}/X_0 \sim \sigma_D \sqrt {\frac{X_0}{V_D}}$  (Gatheral [2010]).

Given $T=T_\opt$ and $A$, the optimal trading schedule is determined numerically by minimizing the total trading cost $C(T_\opt; \nu)$ with respect to the shape parameter $\nu$:
\be
\nu=\underset{\nu ; \ \nu>1}{\operatorname{argmin}} \left\{ I(T_\opt,\nu)+\frac{1}{2\rho} R^2(T_\opt,\nu) \right\} ~.
\label{eq22}
\ee

The optimal execution time is a function of the trading volume $T_\opt \sim V_D^{-\omega}$ , where $\omega=\frac{\beta}{\beta+1}$.  Thus the uncertainty in volume is translated to the uncertainty in the trading time $T_\opt$ and, correspondingly, to the uncertainty bands of the trading schedule $Y_\text{tgt,min,max}(t)=X_0 \left (1-t / T_\text{tgt,min,max} \right)^\nu$.

The distribution of the trading volume $V_D$ can be approximated by the log-normal distribution $Log \textit{N}(\mu_Z,\sigma_Z^2)$.
Using the standard property of the log-normal random variables $V_D^{-\omega} \sim Log \textit{N}(-\omega \mu_Z , \omega^2 \sigma_Z^2)$  (Aitchison and  Brown [1957]), the mean $\mu(V_D^{-\omega})$ and the variance  $\sigma^2(V_D^{-\omega})$ of log-normal random variable $V_D^{-\omega}$ have the form
\begin{eqnarray*}
\mu(V_D^{-\omega}) & = & \exp\left(-\omega \mu_Z+\frac{(\omega\sigma_Z)^2}{2} \right) ~, \\
\sigma^2(V_D^{-\omega}) & = & \left[ \exp\left( \omega \sigma_Z \right)^2 - 1 \right] \times \exp\left(-2 \omega \mu_Z+{(\omega \sigma_Z)^2}\right) ~.
\end{eqnarray*}
The durations $T_\tgt$, $T_\text{max}$, and  $T_\text{min}$ can be derived using
\be
T_\tgt = c \mu(V_D^{-\omega}) ~, \ \
T_\text{min} = c (\mu(V_D^{-\omega}) - \eta \sigma (V_D^{-\omega})) ~, \ \
T_\text{max} = c (\mu(V_D^{-\omega})+\eta \sigma (V_D^{-\omega})) ~,
\ee
where $c=X_0^{\omega} \left(6 \beta I_0 /A \right)^{\frac{1}{\beta+1}}$ and $\eta$ is a discretion parameter.  The uncertainty of the future volatility can be incorporated in a similar way.

\begin{figure}[h]
\centering
\includegraphics[width=3in]{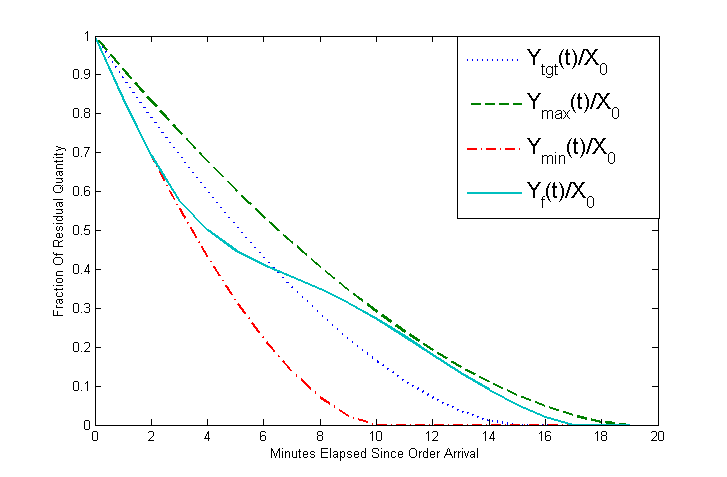}
\caption{Example of \ais\ trading trajectory with uncertainty bands} \label{fig:States2}
\end{figure}

We compute a numerical example for the purpose of exploring the quantitative properties of the proposed approach.  Consider an order to buy $X_0=1M$ shares of a stock currently priced at $P_0=24.7$.  The average daily volume $V_D=70M$, the daily volatility $\sigma_D=0.0113$, $\beta=0.5$, and $I_0 = 0.1$.  With aggressiveness $A=5$ the optimal duration in volume time is $T_\opt=0.037$ ($\approx 15$ minutes), and optimal average participation rate $p_\opt= 38\%$. The optimal shape parameter obtained numerically is  $\nu_\opt=1.65$.
The respective mean and standard deviation of the log normal daily volume are $\mu_Z = 18$ and $\sigma_Z = 0.4$. Thus the inverted square root of volume is parameterized with $\mu(V_D^{-0.5})=1.3\times 10^{-4}$ and $\sigma(V_D^{-0.5})=0.4 \times 10^{-4}$. The target trading time is $T_\tgt=T_\opt=0.037$, the minimum trading time is $T_\text{min} = 0.025$ ($\approx 10$ minutes), and the maximum trading time is $T_\text{max} = 0.049$ ($\approx 19$ minutes).  Examples of trading trajectories with $\eta=1$ calculated in this model are shown in Figure \ref{fig:States2}.

\section*{Implementation: Discretized-Time Approach}

The continuous-time formulation of schedule-based strategies described above is stateless, in the sense that it is only necessary to know the bands at the current time $t$.  In an  alternative scheme for trading within uncertainty bands, the trading interval is divided into contiguous subintervals, or {\em bins}.  The underlying time coordinates are any of clock time, trade time, or volume time. The strategy allocates shares to be executed within the current bin (and, optionally, future bins).  Attention must be paid to the cost of cleaning up shares that are not executed within the current bin.  As shown in (Jeria, Schouwenaars, and Sofianos [2009]), clean-up costs can significantly exceed the spread capture savings of a particular strategy within the subinterval.  We expect that a similar pattern holds for any strategy designed under the same principle.

To minimize clean-up costs we propose the following approach. Let $\tau$ be the time variable in the chosen coordinate system and assume that the interval $[\tau_0, \tau_1]$ is divided into $N$ bins of uniform width $\delta \tau = (\tau_1 - \tau_0) / N$ each.  Let $T_k$ be the width of the $k^{th}$ bin in clock time.  We assume the existence of an opportunistic, short-duration (``tactical'') strategy that supports an order duration and minimum fill amount.

The strategy begins trading at the beginning of the first bin ($\tau = \tau_0$).  The range of shares that should be completed as of the end of the first bin is $\left[X_\text{min}(\tau_0 + \delta\tau), X_\text{max}(\tau_0 + \delta\tau)\right]$.  At this time, this is the only information about the trading schedule that is needed.  The strategy routes $X_\text{max}(\tau_0 + \delta\tau)$ shares to the tactical strategy with a duration of $T_1$ and minimum fill amount $X_\text{min}(\tau_0 + \delta\tau)$.

At the start of the second bin (end of the first bin) the strategy has filled $X_f(\tau_0 + \delta\tau)$ shares, where $X_\text{min}(\tau_0 + \delta\tau) \le X_f(\tau_0 + \delta\tau) \le X_\text{max}(\tau_0 + \delta\tau)$.  Based on this outcome and the recent and current market conditions, the next range $\left[X_\text{min}(\tau_0 + 2\delta\tau), X_\text{max}(\tau_0 + 2\delta\tau)\right]$ is calculated.  The schedule-based strategy routes $X_\text{max}(\tau_0+2\delta\tau) - X_f(\tau_0 + \delta\tau)$ shares to the tactical strategy with a duration $T_2$ and minimum fill amount $X_\text{min}(\tau_0+2\delta\tau) - X_f(\tau_0+\delta\tau)$.  This cycle is repeated until the order is completed or the end time is reached.

For example, consider VWAP and select volume time as our time coordinate, so that the trading trajectory is linear.  At the start of trading (first bin), we set $X_\text{min}(\tau_0+\delta\tau) = 0$ and $X_\text{max}(\tau_0+\delta\tau) = X_0 / N$.  We route $X_0/N$ shares to the tactical strategy for duration $T_1$ and no minimum fill amount, so that the tactic has full discretion to execute any amount up to $X_0/N$.

At time $\tau = \tau_0+\delta\tau$, we require the unexecuted shares $X_0/N - X_f(\tau_0+\delta\tau)$ to be filled with certainty in the next interval.  Thus we route $2 X_0/N - X_f(\tau_0+\delta\tau)$ shares to the tactical strategy for duration $T_2$ and minimum fill $X_0/N - X_f(\tau_0+\delta\tau)$.  We continue this process bin by bin, carrying forward any shortfall to the aggressive component of the next slice.

\section*{Conclusion}

We have presented a practical design of sell-side schedule-based trading strategies. This design allows simple implementation of the popular VWAP, Participation, and Implementation Shortfall strategies in both continuous and discrete-time approaches.  It cleanly separates high-level scheduling from low-level execution tactics.  The partition of the schedule into aggressive, passive, opportunistic, and dark shares is dictated by the filled shares position relative to the bands and the allocation among their respective execution tactics is de-coupled from the schedule generation.  The band separation gives the strategy discretion to wait and exploit profitable price and liquidity patterns and the framework easily incorporates an $\alpha$ model for this purpose.

\section*{Acknowledgements}
The authors thank Vacslav Glukhov and David Fellah for helpful discussions and suggestions.
\section*{References}

Aitchison J., and J.A.C. Brown,
``The Lognormal Distribution'', Cambridge University Press, 1957.

Almgren, R., and N. Chriss,
``Optimal Execution of Portfolio Transactions'', J. Risk, vol. 3, n.2, 2000, pp. 5-39.

Altunata, S., D. Rakhlin, and H. Waelbroeck,
``Adverse Selection vs. Opportunistic Savings in Dark Aggregators'', Journal of Trading, vol. 5, 2010, pp.16-28.

Bouchaud, J.-P., J. D. Farmer, and F. Lillo,
``How markets slowly digest changes in supply and demand'', http://arxiv.org/abs/0809.0822.

Bouchaud, J.-P., and M. Potters,
``Theory of Financial Risk and Derivative Pricing: From Statistical Physics to Risk Management'', Cambridge University Press, 2004.

Gatheral J.,
``No-dynamic-Arbitrage and Market Impact'', Quantitative Finance, vol.10, n.7, 2010, pp.749-759.

Glukhov S.,
``Optimal Trading in the Presence of Non-Displayed Liquidity'', Journal of Trading, vol. 2, 2007, pp.30-37.

Jeria D., T. Schouwenaars, and G. Sofianos,
``The all-in cost of passive limit orders'', Street Smart, Issue 38, Goldman Sachs, 2009.

Kissell, R., and M. Glantz,
``Optimal Trading Strategies: Quantitative Approaches for Managing Market Impact and Trading Risk'', AMACOM, 2003.

Mazur S.,
``On U-Curves and Uncertainty Bands'', Liquidnet Technical Report, 2011.

Zaitsev, S., A. Zaitsev, A. Leonidov, and V. Trainin,
``Market mill dependence patter in the stock market: Multiscale conditional dynamic'', Physica A, vol. 388, n. 21, 2009, pp. 4624-4634.


\end{document}